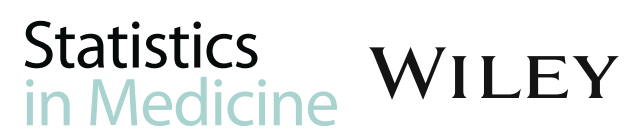

# A two-step log-linear procedure for graphical representation and inference of associations in cross-classified data for disease diagnosis

**J. Fernando Vera** | **José A. Roldán-Nofuentes**

Department of Statistics and O.R., University of Granada, Granada, Spain

**Correspondence**
J. Fernando Vera, Department of Statistics and O.R., University of Granada, Granada, Spain.
Email: jfvera@ugr.es

**Funding information**
ERDF/Ministry of Economic Transformation, Industry, knowledge and Universities of Andalucía, Grant/Award Number: B-CTS-184-UGR20; MCIN/AEI/10.13039/501100011033 and by "ERDF A way of making Europe", Grant/Award Number: PID2021-126095NB-100; Ministry of Science and Innovation-State Research Agency/10.13039/501100011033/Spain, by "ERDF A way of making Europe", Grant/Award Number: RTI2018-099723-B-I00

Biometrical sciences and disease diagnosis in particular, are often concerned with the analysis of associations for cross-classified data, for which distance association models give us a graphical interpretation for non-sparse matrices with a low number of categories. In this framework, usually binary exploratory and response variables are present, with analysis based on individual profiles being of great interest. For saturated models, we show the usual linear relationship for log-linear models is preserved in full dimension for the distance association parameterization. This enables a two-step procedure to facilitate the analysis and the interpretation of associations in terms of unfolding after the overall and main effects are removed. The proposed procedure can deal with cross-classified data for profiles by binary variables, and it is easy to implement using traditional statistical software. For disease diagnosis, the problems of a degenerate solution in the unfolding representation, and that of determining significant differences between the profile locations are addressed. A hypothesis test of independence based on odds ratio is considered. Furthermore, a procedure is proposed to determine the causes of the significance of the test, avoiding the problem of error propagation. The equivalence between a test for equality of odds ratio pairs and the test for equality of location for two profiles in the unfolding representation in the disease diagnosis is shown. The results have been applied to a real example on the diagnosis of coronary disease, relating the odds ratios with performance parameters of the diagnostic test.

## 1 | INTRODUCTION

In general, the most data analysis problems in medical research and clinical decisions involve the analysis of associations between a binary response variable and a set of response (or explanatory) categorical variables; that is, the traditional log-linear (or multivariate logistic regression) problem. In particular, this is usual in the diagnosis of a disease, which is a fundamental phase in clinical practice, prior to the treatment and prognostic phases. The diagnosis of a disease is made through the application of diagnostic tests and also considering the symptomatology of the patient and the covariates related to the disease (eg, gender, the presence of risk factors, family history, etc.). For example, an adult male with fatigue may require an electrocardiogram for the diagnosis of coronary heart disease. However, if the male is young, then fatigue may be due to other causes (eg, anemia or diabetes), and therefore the patient will initially require a laboratory test. Therefore, knowing covariates related to the disease is also of great help for the correct diagnosis of the disease.

In many situations, a preliminary step is taken to decide the most influential variables in a set, but often this selection may not be so clear. Instead of this kind of variable-oriented approach, a person-oriented approach to data analysis might be preferred.[1] Any combination of the categories of the variables is called a profile. In the most general case, the data consist of a profile-by-response contingency table in which the profiles represent a single categorical response variable. In this situation, the analysis of associations between profiles of individuals based on diagnostic test together with covariates (rows), and having or not having a disease (columns), is of great interest. In the clinical setting, this allows greater control over the profiles with the highest risk, which can favor personalized treatment and improve health outcomes.

Distance association (DA) models enable the representation of associations for a saturated log-linear model in terms of squared Euclidean distances.[2-4] It has been shown that the DA model produces the same expected frequencies than the RC(M) model,[5] although the graphical interpretation of the associations is easier in the DA model in terms of points in a Euclidean space.[2] This is performed by considering a suitable parameterization of the association term that, together with a transformation for the identification of the parameters in the model, preserves the log-linear relation. In low dimensions, the distance association model is an approximation to the traditional log-linear model and, in general, it is more efficient than a two-step procedure that first performs traditional log-linear analysis and then represents the associations in terms of Euclidean distances. In a combined procedure of clustering and distance association model to deal with sparse data sets, the superiority of a simultaneous strategy compared to a two-step procedure has been explicitly shown by Vera, de Rooij and Heiser[3] for the latent class distance association model and by Vera and de Rooij[4] for the latent block distance association model.

In DA models, associations are analyzed relative to an unconditional unfolding procedure after removing the effect of the overall mean, and the main row and column effects, in a log-linear framework. However, when a binary variable is considered as a response, while the profiles are a categorical explanatory variable, a row conditional representation may also be appropriate. For example, when one of the two variables is binary, a simple and easy-to-interpret conditional representation can be made when the associations are analyzed directly from the observed frequencies (without removing the overall and main effects) using the procedure proposed by Heiser.[6] This exploratory procedure relates the position of the points (centroids and vertexes) in a simplex with the distances between them and the observed proportions.

An important feature in DA models, and in particular for the diagnosis of a disease, is to statistically determine significant differences between the profile locations in the unfolding representation. This is a very important issue in this context and one that is not traditionally covered by unfolding procedures. In general, unfolding is based on the relationships between the elements of two sets to be represented, in this case the profiles of the row variable and the two column categories. However, unfolding does not usually address the relationships within the categories of each variable separately, and this is also true of the distance association model. Stability analysis of point locations has been used to determine confidence regions in multidimensional scaling,[7,8] but an inference-based procedure in this respect in unfolding, and in particular in the distance association model, would be desirable.

In general, when two qualitative variables are studied in a $I \times J$ table, the independence between them is usually studied by the classical Fisher's exact test or the chi-squared test.[9] In a distance association model, it is also of interest to study the independence between an explanatory or response variable (such as the presence or absence of a disease) and profiles (eg, based on diagnostic tests and covariates), in particular, in terms of estimated expected frequencies. Testing independence is equivalent to testing that all odds ratios between profiles are equal to one. This will be of special interest when the representation of associations is one-dimensional, as occurs in the diagnosis of a disease, given the relationship between odds ratio and distances. It is important to note that studying the independence for all the profiles across all combinations of individual hypothesis tests about the odds ratio at the $\alpha$ error (or equivalently through the respective confidence intervals at the level $1 - \alpha$), can lead to erroneous conclusions due to the propagation of the $\alpha$ error. In a distance association model for the disease diagnostic, this can be related to the location of the points in the unfolding representation which highly complements the graphic interpretation.

In this article, we show that, in full dimension, a distance-based parameterization of the association terms leads to an equivalent linear decomposition of the expected frequencies to that given by the traditional log-linear model, which means the distance association model in full dimension is a log-linear model with a special interpretation of the parameters. This result is particularly relevant when the variable of interest is binary, as is usual in the diagnosis of a disease, or is determined by a categorical variable with at most four categories, (eg, non-diseased, patients in phase 1, patients in phase 2 and patients in phase 3), which allows the associations to be visualized graphically in full dimension. In this situation, a two-step procedure is proposed that first estimates a log-linear model, for example, using any statistical software familiar to medical researchers, and then represents the estimated associations in full dimension using the parameterization of a distance association model. From a computational point of view, the expected frequencies in any saturated log-linear model coincide with the observed frequencies, so only the DA parameterization of the observed frequencies is needed for the graphical representation, which makes it even easier to use. However, since the model is a true log-linear model, it allows the use of hypothesis tests and confidence intervals for the validation of the traditional log-linear model, while also enabling clinical researchers to take decisions easily based on an exact graphical representation of the associations in low dimension.

Moreover, based on the relationship between the odds ratio and the Euclidean distances in the unfolding representation, we propose an inferential procedure to decide about a degenerate solution[10] in the DA model, and also about significant differences between two locations in the unfolding representation in terms of the odds ratio. In this context, when the profiles correspond to results of binary diagnostic tests and categorical covariates, the odds that a profile has the disease can be expressed in terms of measures of accuracy of the diagnostic tests in the different patterns of the covariates. In the same way, the odds ratios between each two profiles can be expressed. Thus, if the profiles are obtained from a diagnostic test and two binary covariates, then the odds that a profile has the disease can be written in terms of the post-test odds of the diagnostic test, and the odds ratio between two profiles can be expressed in terms of the likelihood ratios (or post-test odds) of the diagnostic test in the different patterns of the covariates. Therefore, the Euclidean distances in the unfolding representation are related to measures of the quality of the binary diagnostic test in the different patterns of covariates.

The rest of the article is organized as follows. In the next section, we formulate the two-step log-linear procedure for estimating and representing associations for cross-classified data, and the equivalence of the log-linear model to the full-dimensional distance association model is shown in terms of their parameterizations. Furthermore, for the case of association with a binary variable, the relationship between independence and equality of location of the points in the unfolding representation is shown in terms of odds ratio. In particular, the relationship between a degenerate solution and independence is also shown. Section 3 introduces the independence test in terms of odds ratios for a general distance association model. Additionally, in the case of the association with a binary variable, the independence test is studied in terms of odds ratio, together with its properties, and a method is proposed to study the causes of significance when the independence test is significant. The results are also specified for the situation in which a diagnostic test and two binary covariates are available, relating the odds ratio between two profiles with measures of the accuracy of the binary diagnostic test. In Section 4, a Monte Carlo experiment is performed to study the asymptotic behavior of the hypothesis test introduced above, using eight profiles. In Section 5, the results have been applied to the diagnosis of coronary heart disease. In the final section, we summarize the main conclusion drawn.

## 2 | REPRESENTATION OF ASSOCIATIONS BASED ON DISTANCES

Let us denote by $\mathbf{F} = (f_{ij})$ an $I \times J$ contingency table that collects the counts of combinations of row and column categories, which can represent profiles of variables. We will consider here the particular situation in which the rows represent profiles or categories of a response variable and the columns represent possible stages or characteristics of a disease. Let us define the $I \times M$ matrix $\mathbf{X}$ and the $J \times M$ matrix $\mathbf{Y}$, whose row vectors $\mathbf{x}_i$, $i = 1, \ldots, I$ and $\mathbf{y}_j$, $j = 1, \ldots, J$ are the coordinates of the row and column categories of $\mathbf{F}$, respectively, in dimension $M$.

We use the well-known equivalence of the multinomial and Poisson distribution,[9,11] and under the usual Poisson sampling model, counts are considered as independent random variables,[9] and denoting by $\mu_{ij}$ the expected values, the log-likelihood is given by

$$\log(L) = \sum_{i=1}^{I}\sum_{j=1}^{J} f_{ij}\log(\mu_{ij}) - \sum_{i=1}^{I}\sum_{j=1}^{J}\mu_{ij}. \tag{1}$$

In the usual multiplicative form in a log-linear framework for a two-way cross-classification, the expected frequencies can be written as follows:

$$\mu_{ij} = \mu\alpha_i\beta_j\theta_{ij}, \tag{2}$$

where $\mu$ is the overall scale parameter, $\alpha_i$ is the row effect parameter, $\beta_k$ is the column effect parameter, and $\theta_{ij}$ is the interaction effect. In the distance association model is assumed the association parameter $\theta_{ij}$ can be expressed in terms of the Euclidean distances,

$$d^2(\mathbf{x}_i, \mathbf{y}_j) = \sum_{m=1}^{M} (x_{im} - y_{jm})^2$$

using the exponential form $\theta_{ij} = \exp\left(-d_{ij}^2\right)$. Thus, the greater the association between a row and column categories, the smaller the distance between their corresponding points. Taking the logarithm of (2) we can express the model in log-linear terms (focusing on the squared distances) as follows,

$$\log(\mu_{ij}) = \lambda + \lambda_i + \lambda_j - d_{ij}^2(\mathbf{x}_i, \mathbf{y}_j), \tag{3}$$

where $\log(\theta_{ij}) = -d_{ij}^2$ (note that this model is the distance association model, which in low dimension is log-quadratic in the configuration parameters, but in full dimension is log-linear as shown in Appendix A). As usual, for identification purposes, the mean of the row effects and the mean of the column effects will be assumed to be zero. To this end, the following parameterization of the expected frequencies[2] can be considered.

## 2.1 | Equivalence between log-linear and distance association models in full dimension

Given the parameter estimated values from (3) by the usual saturated log-linear (see Appendix A) or by the distance association model in full dimension, let us denote by $\widehat{\boldsymbol{\mu}} = (\widehat{\mu}_{ij})$ the $I \times J$ matrix of the estimated expected frequencies ($\widehat{\boldsymbol{\mu}} = \boldsymbol{\mu}$, since both are saturated models). To obtain an identified solution, the parameters are expressed as a function of singular values and singular vectors,[2] since the singular value decomposition is unique and is characterized by $M(M+2)$ constraints, with $M = \min(I, J) - 1$ the maximum dimension.[10]

Let us denote by $\mathbf{G} = (g_{ij})$, the matrix of entries $g_{ij} = \log(\widehat{\mu}_{ij})$, and denote by $\overline{\overline{g}}$, the global mean of $\mathbf{G}$, and by $\overline{g}_i$ and $\overline{g}_j$ the marginal means for the $i$th row and the $j$th column of $\mathbf{G}$, respectively. Then, we define $\widetilde{\lambda} = \overline{\overline{g}}$, $\widetilde{\lambda}_i^R = \overline{g}_i - \overline{\overline{g}}$, $\widetilde{\lambda}_j^C = \overline{g}_j - \overline{\overline{g}}$, and $\boldsymbol{\Delta}$ the matrix of entries $\delta_{ij} = g_{ij} - \widetilde{\lambda} - \widetilde{\lambda}_i^R - \widetilde{\lambda}_j^C$. From the singular value decomposition of $\boldsymbol{\Delta} = \mathbf{U}\boldsymbol{\Gamma}\boldsymbol{\Lambda}'$, we can estimate $\mathbf{X}$ and $\mathbf{Y}$, both in dimension M, such that $\mathbf{X}\sqrt{2} = \mathbf{U}\boldsymbol{\Gamma}^{1/2}$ and $\mathbf{Y}\sqrt{2} = \boldsymbol{\Gamma}^{1/2}\boldsymbol{\Lambda}'$. Then, denoting by $d_{x,i} = \sum_m x_{im}^2$, and $d_{y,j} = \sum_m y_{jm}^2$, identified parameters are obtained:

$$\dot{\lambda}_i^R = \widetilde{\lambda}_i^R + d_{x,i}, \tag{4}$$

$$\dot{\lambda}_j^C = \widetilde{\lambda}_j^C + d_{y,j}, \tag{5}$$

$$\lambda = \widetilde{\lambda} + \frac{1}{I}\sum_{i=1}^{I}\dot{\lambda}_i^R + \frac{1}{J}\sum_{j=1}^{J}\dot{\lambda}_j^C, \tag{6}$$

$$\lambda_i^R = \dot{\lambda}_i^R - \frac{1}{I}\sum_{i=1}^{R}\dot{\lambda}_i^R, \tag{7}$$

$$\lambda_j^C = \dot{\lambda}_j^C - \frac{1}{J}\sum_{j=1}^{J}\dot{\lambda}_j^C. \tag{8}$$

The mean of the values of $\lambda_i^R$, $i = 1, \dots, I$ and of $\lambda_j^C$, $j = 1, \dots, J$ is equal to zero, and $g_{ij} = \lambda + \lambda_i^R + \lambda_j^C - d_{ij}^2(\mathbf{x}_i, \mathbf{y}_j)$. Then, after this parameterization, the model is characterized by $2 + M(M+2)$ further constraints.[2] Thus, in full dimension $M$, two configurations $\mathbf{X}$ and $\mathbf{Y}$ can be estimated such that (3) holds. This results in a distance association model which is equivalent to the corresponding log-linear model, and therefore, a two-step procedure for the parameter estimation is allowed. The parameters of the log-linear model can be first estimated using any statistical software for inferential purpose, or simply consider the expected (observed) frequencies, and then the above parameterization can be used to represent associations in full dimension.

## 2.2 | Odds ratio and distances

The log odds of a response $j$ against a response $j'$ for a given class $i$ are given by

$$log\left(\frac{\mu_{ij}}{\mu_{ij'}}\right) = \log(\beta_j) - \log(\beta_{j'}) - d_{ij}^2 + d_{ij'}^2.$$

The odds are a function of both the main effect parameters and the distances. Concerning the distances, the odds are in favor of the closest category. Concerning the main effects, the odds are in favor of the category with the largest $\beta$ value. The odds ratio can be defined in terms of squared distances as[2]

$$O_{ii'jj'} = \frac{\mu_{ij} \times \mu_{i'j'}}{\mu_{ij'} \times \mu_{i'j}} = \exp\left(-d_{ij}^2 - d_{i'j'}^2 + d_{ij'}^2 + d_{i'j}^2\right), \tag{9}$$

and if the row category $I$ and the column category $J$ are set as the reference, then the usual set of local odds ratio, denoted by $O_{ij} = O_{iIjJ}$, are given in terms of distances as $O_{ij} = \exp\left(-d_{ij}^2 - d_{IJ}^2 + d_{iJ}^2 + d_{Ij}^2\right)$. Then, for a fixed column $j$, $O_{ij} = O_{i'j}$ if and only if $\left(d_{iJ}^2 - d_{ij}^2\right) = \left(d_{i'J}^2 - d_{i'j}^2\right)$.

When the cross-classified data set involves a binary variable, the baseline category $J$ represents one of the two choices, and the unfolding representation is in one dimension. Then, denoting by $J$ and $j$ the only two column categories in the table, it follows that for any $i, i' = 1, \ldots I$ row categories,

$$O_{ij} = O_{i'j} \Leftrightarrow d_{iJ} - d_{ij} = d_{i'J} - d_{i'j}, \tag{10}$$

which only occurs when both categories $i$ and $i'$ are represented by the same point in the unfolding configuration (see Appendix B). Hence, from a statistical point of view, a hypothesis test contrasting the equality of the corresponding odds ratios would allow us to decide whether the position of two nearby points can be considered to be significantly different in terms of association. When the response variable is binary, $O_{iIjJ} = O_{i'IjJ}$ if and only if $O_{ii'jJ} = 1$. Therefore, the hypothesis that two row profiles do not differ from each other with respect to their association with the response variable can be formulated in terms of the independence test by considering the null hypothesis: $O_{ii'jJ} = 1$.

# 3 | TESTING DIFFERENCES IN DISTANCE ASSOCIATIONS BETWEEN PATIENT PROFILES

The study of patient profiles and their association with a certain disease is of great interest in Clinical Medicine and Preventive Medicine. Here we will focus on patient profiles defined from qualitative variables, such as the results of binary diagnostic tests and disease-related covariates. Likewise, we will also consider that the disease status of each patient (present disease or absent disease) is known by applying a gold standard (GS), which is the medical test that allows to determine whether or not a patient has the disease. In this context, consider a random sample of $n$ individuals, to whom $T$ binary diagnostic tests (BDTs) are applied, and $K$ qualitative covariates $A_1, \ldots, A_K$ are observed in all individuals. Each covariate can take values $a_1, a_2, \ldots$ Let $T_t$ be the binary random variable that models the result of $t$-ht BDT, with $t = 1, \ldots, T$, such that $T_t = 1$ when the test is positive and $T_t = 2$ when it is negative, and $GS$ models the result of the gold standard ($GS = 1$ if the individual has the disease and $GS = 2$ if he does not have it). Let $a_{a_1,\ldots,a_K,t_1,\ldots,t_T}$ and $b_{a_1,\ldots,a_K,t_1,\ldots,t_T}$ be the number of diseased and non-diseased individuals, respectively, in which $A_1 = a_1, \ldots, A_K = a_K$ and $T_1 = t_1, \ldots, T_T = t_T$, with $t_t = 1, 2$ and $t = 1, \ldots, T$. The observed frequencies are the realization of a multinomial distribution with

probabilities

$$p_{a_1,\ldots,a_K,t_1,\ldots,t_T} = P(GS = 1, A_1 = a_1, \ldots, A_K = a_K, T_1 = \mathrm{t}_1, \ldots, T_T = \mathrm{t}_T)$$

and

$$q_{a_1,\ldots,a_K,t_1,\ldots,t_T} = P(GS = 2, A_1 = a_1, \ldots, A_K = a_K, T_1 = \mathrm{t}_1, \ldots, T_T = \mathrm{t}_T),$$

such that

$$\sum p_{a_1,\ldots,a_K,t_1,\ldots,t_T} + \sum q_{a_1,\ldots,a_K,t_1,\ldots,t_T} = 1.$$

The maximum likelihood estimators of these probabilities are

$$\hat{p}_{a_1,\ldots,a_K,t_1,\ldots,\mathrm{t}_T} = \frac{a_{a_1,\ldots,a_K,t_1,\ldots,\mathrm{t}_T}}{n}$$

and

$$\hat{q}_{a_1,\ldots,a_K,t_1,\ldots,\mathrm{t}_T} = \frac{b_{a_1,\ldots,a_K,t_1,\ldots,\mathrm{t}_T}}{n}.$$

Let $\boldsymbol{\pi}$ be a vector of dimension $2I$ whose components are the above probabilities, where $I$ is the number of profiles. Applying the multivariate central limit theorem it is verified that

$$\sqrt{n}(\hat{\boldsymbol{\pi}} - \boldsymbol{\pi}) \underset{n\to\infty}{\longrightarrow} N_{2I}\left(\mathbf{0}, \sum_{\boldsymbol{\pi}}\right).$$

Since the sample of size $n$ is the realization of a multinomial distribution, the variance-covariance matrix $\sum_{\boldsymbol{\pi}}$ is estimated as follows:

$$\widehat{\sum}_{\hat{\boldsymbol{\pi}}} = \left\{\mathrm{diag}(\hat{\boldsymbol{\pi}}) - \hat{\boldsymbol{\pi}}\hat{\boldsymbol{\pi}}^T\right\}/n.$$

Assuming $GS = 1$ the baseline category, let $O_{ii'} = O_{ii'21}$ be the odds ratio between profile $i = (a_1, \ldots, a_K, t_1, \ldots, \mathrm{t}_T)$ and profile $i' = \left(a'_1, \ldots, a'_K, t'_1, \ldots, \mathrm{t}'_T\right)$, with $i, i' = 1, \ldots, I$ and $i \neq i'$. If $O_{ii'} = 1$, then the odds of having the disease is the same for both profiles. It is obvious that $O_{ii} = 1$ and that $O_{ii'} = O_{i'i}^{-1}$. In terms of the probabilities of the vector $\boldsymbol{\pi}$, the odds ratio between $i$ and profile $i'$ is written as follows:

$$O_{ii'} = \frac{p_{a_1,\ldots,a_K,t_1,\ldots,\mathrm{t}_T}/q_{a_1,\ldots,a_K,t_1,\ldots,\mathrm{t}_T}}{p_{a'_1,\ldots,a'_K,t'_1,\ldots,\mathrm{t}'_T}/q_{a'_1,\ldots,a'_K,t'_1,\ldots,\mathrm{t}'_T}},,$$

and its estimator is

$$\hat{O}_{ii'} = \frac{a_{a_1,\ldots,a_K,t_1,\ldots,\mathrm{t}_T}/b_{a_1,\ldots,a_K,t_1,\ldots,\mathrm{t}_T}}{a_{a'_1,\ldots,a'_K,t'_1,\ldots,\mathrm{t}'_T}/b_{a'_1,\ldots,a'_K,t'_1,\ldots,\mathrm{t}'_T}}.$$

In this situation, it is of interest to study whether or not the profiles are associated with the disease, and for this reason the following hypothesis test is studied.

## 3.1 | Testing independence and degenerate solutions in DA

In general, the distance association model in low dimension is not equivalent to the related log-linear model, except for $M = \min(I, J) - 1$. In addition, the DA model represents associations between row and column categories, after overall, row and column effects in the cross-classification dataset are removed. Hence, a procedure for testing independence in terms of the estimated distances is of great interest, in particular when the DA model is estimated in a dimension lower than $M$. As noted above, testing the independence between the $I$ profiles and the disease is equivalent to testing that all odds ratios are equal to 1, and that therefore all profiles have the same odds of having the disease. In this situation, independence is equivalent to the fact that the location of all profiles is the same in the unfolding representation, that is, that the solution is degenerate.[10,12] Here, we focus on the situation in which the associations are represented in full dimension, in which case the log-linear and DA models produce the same expected frequencies, which coincide with the observed frequencies as they are saturated models. Although global independence can be tested using any classical test, or by comparing models in a log-linear framework; here, we focus on odds ratio tests and its properties, in line with the identification of degenerate solutions. Having set a profile $i$, $i \in \{1, \ldots, I\}$, $O_{ii'} = O_{i''i'}/O_{i''i}$ with $i' \neq i''$, and an independence test, or global hypothesis test to compare that all odds ratios are equal to 1, is defined as follows:

$$
\begin{aligned}
&H_0 : O_{ii'} = 1, i' = 1, \ldots, I, i \neq i', \\
&H_1 : \text{al least one odds ratio is different from } 1. 
\end{aligned} \tag{11}
$$

In this hypothesis test, only $I - 1$ odds ratios are involved, and the null hypothesis means that all odds ratios are equal to 1, regardless of the baseline profile. In this paper, we solve this contrast by applying transformations on the odds ratio, specifically the natural logarithm, which is a transformation widely used to compare and estimate parameters and whose application to the case of an odds ratio in a $2 \times 2$ table is widely known. Therefore, the hypothesis test (11) is equivalent to the test

$$
\begin{aligned}
&H_0 : U_{ii'} = 0, \ i' = 1, \ldots, I, \ i' \neq i, \\
&H_1 : \text{al least one } U_{ii'} \text{ is different from } 0,
\end{aligned} \tag{12}
$$

where $U_{ii'} = \log(O_{ii'})$. Let $\mathbf{U_i} = (U_{i1}, \ldots, U_{iI})^T$, that is, $\mathbf{U_i}$ is a vector of dimension $I - 1$ whose components are $U_{ii'}$, with $i, i' = 1, \ldots, I$ and $i' \neq i$. Applying the multivariate central limit theorem it is verified that

$$\sqrt{n}\left(\widehat{\mathbf{U}}_i - \mathbf{U}_i\right) \xrightarrow[n \to \infty]{} N_{I-1}\left(\mathbf{0}, \boldsymbol{\Sigma}_{\mathbf{U}_i}\right).$$

As $\mathbf{U}_i$ is a function of the probabilities of $\boldsymbol{\pi}$, the estimated asymptotic variance of $\sum_{\mathbf{U}_i}$ is obtained applying the delta method, that is,

$$\widehat{\sum}_{\widehat{\mathbf{U}}_\mathbf{i}} = \left(\frac{\partial \mathbf{U}_i}{\partial \boldsymbol{\pi}}\right)_{\boldsymbol{\pi}=\widehat{\boldsymbol{\pi}}} \widehat{\sum}_{\widehat{\boldsymbol{\pi}}} \left(\frac{\partial \mathbf{U}_i}{\partial \boldsymbol{\pi}}\right)^T_{\boldsymbol{\pi}=\widehat{\boldsymbol{\pi}}}.$$

Then, the test statistic

$$Q_i = \widehat{\mathbf{U}}_i^T \widehat{\sum}_{\widehat{\mathbf{U}}_i}^{-1} \widehat{\mathbf{U}}_i$$

is distributed according to Hotelling's $T$-squared distribution with a dimension $I - 1$ and $n$ degrees of freedom, where $I - 1$ is the dimension of the vector $\widehat{\mathbf{U}}_i$. When $n$ is large, the statistic $Q_i$ is distributed according to a central chi-squared distribution with $I - 1$ degrees of freedom when the null hypothesis is true, that is,

$$Q_i = \widehat{\mathbf{U}}_i^T \widehat{\sum}_{\widehat{\mathbf{U}}_i}^{-1} \widehat{\mathbf{U}}_i \xrightarrow[n \to \infty]{} \chi^2_{I-1}.$$

Therefore, having set a profile we will obtain a value for the test statistic of the global test and the corresponding $P$-value. It can be shown (see Appendix C) that using the transformation of the natural logarithm, the test statistic for the global test is the same regardless of the baseline profile set, that is,

$$Q_i = Q_{i'} = Q, \ i, i' = 1, \ldots, I, \ i' \neq i.$$

Setting an $\alpha$ error, if the $P$-value of the hypothesis test (12) is higher than $\alpha$ then we do not reject that all of the profiles have the same odds of having the disease, that is, the unfolding configuration is degenerate. For a $P$-value $\leq \alpha$, we accept that at least one odds ratio is different from 1, which means that at least one profile has odds of having the disease different to the rest of the profiles. Faced with this situation, it is necessary to investigate the causes of the significance of the test, for which the usual steps are:

1. Solving the individual hypothesis tests on the odds ratio between two profiles. This is of special interest in this framework, since as shown above in Section 2, this is equivalent to testing whether the distance between the locations of two row profiles differs from zero in the DA model representation, that is, the localization of the two profiles is the same in the unfolding representation. Hence, we propose to statistically determine if there are any differences between the associations of two row profiles with respect to having the disease or not. To this end, we study the test
$$H_0 : U_{ii'} = 0 \text{ vs. } H_1 : U_{ii'} \neq 0,$$
with $i < i'$ and $i, i' = 1, \dots, I$. A Wald test statistic for this hypothesis test is[9]
$$z_{ii'} = \frac{\hat{U}_{ii'}}{\sqrt{\hat{V}ar\left(\hat{U}_{ii'}\right)}},$$
which is distributed according to a normal standard distribution when $n$ is large, where $\hat{V}ar\left(\hat{U}_{ii'}\right)$ is obtained applying the delta method. This test statistic is the classical one used to test that the log-odds ratio is equal to zero.[9] A confidence interval for the log-odds ratio can be obtained simply by inverting this test statistic (Agresti[9]). However, it is important to take into account that the formulation of hypothesis tests allows us to know the evidence against the null hypothesis. The number of individual hypothesis tests that need to be solved is $I(I-1)/2$.

2. Adjusting the $P$-values to control the $\alpha$ error to adjust the $I(I-1)/2$ $P$-values obtained by solving the individual tests, we propose to use Holm's method,[13] which is easy to apply and is less conservative than the classic Bonferroni method.

The development for the clinical situation in which the diagnosis of the disease is based on the application of a BDT and on the observation of two binary covariates is shown in Appendix D.

## 4 | SIMULATION EXPERIMENTS

Monte Carlo simulation experiments have been carried out to study the type I errors and the powers of the independence test based on the odds ratios ($Q$ statistic), together with the Pearson chi-square test of independence ($\chi^2$ statistic) and likelihood ratio test ($G^2$ statistic).[9] The value of $I = 8$ was considered, such as the situation studied in Appendix D and, for example, when considering two BDTs and a binary covariate, or three BDTs. As usual, the value of 0.5 was added to the entire table when at least one cell has zero frequency (a situation in which none of the three independence tests can be applied). These experiments consisted of the generation of 10 000 multinomial random samples sized $n = \{125, 150, 200, 300, 400, 500, 1000, 2000, 5000\}$ (for $n \leq 100$ it was not possible to generate enough random samples under the necessary conditions[14,15] to apply the classical tests), and whose probabilities were calculated as follows:

1. For $\sum p_{rst}$ were considered the values of $\{0.20, 0.40, 0.60, 0.80\}$, which also set the values of $\sum q_{rst}$ (since $\sum q_{rst} = 1 - \sum p_{rst}$).
2. For $p_{111}$ were considered the values of $\left\{0.05, \dots, \sum p_{rst} - 0.05\right\}$ and for $q_{111}$ the values of $\left\{0.05, \dots, \sum q_{rst} - 0.05\right\}$.
3. For simplicity, for $p_{rst}$ and $q_{rst}$, with $(r, s, t) \neq (1, 1, 1)$, were considered the values of $p_{rst} = \frac{\sum p_{rst} - p_{111}}{7}$ and $q_{rst} = \frac{\sum q_{rst} - q_{111}}{7}$, respectively. Therefore, it is considered that $p_{rst}$ ($q_{rst}$) are all equal and so, in each profile the odds ratios are all equal (for $i' > i$), which considerably simplifies the dimension of the problem.

For the type I error, all of the odds ratios are equal to 1, and for the power we considered that in the first profile ($i = (1, 1, 1)$) all of the odds ratios are equal to each other and greater than 1 ($O_{1i'} > 1$ with $i' = 2, \dots, 8$), and that in the

**TABLE 1** Type I errors (in %) of the independence tests.

| | $O_{ii'} = 1$, $\sum p_{ijk} = 0.20$ $\sum q_{ijk} = 0.80$, $p_{111} = 0.05$ $q_{111} = 0.20$ | | | $O_{ii'} = 1$, $\sum p_{ijk} = 0.40$ $\sum q_{ijk} = 0.60$, $p_{111} = 0.10$ $q_{111} = 0.15$ | | |
|---|---|---|---|---|---|---|
| **n** | **Q** | $\chi^2$ | $G^2$ | **Q** | $\chi^2$ | $G^2$ |
| 125 | 2.7 | 8.6 | 12.6 | 2.1 | 4.6 | 6.3 |
| 150 | 2.8 | 7.4 | 10.8 | 2.4 | 4.7 | 6.0 |
| 200 | 2.9 | 5.9 | 7.1 | 3.2 | 4.4 | 5.4 |
| 300 | 3.1 | 5.1 | 6.0 | 3.9 | 4.8 | 5.3 |
| 400 | 3.9 | 4.7 | 5.5 | 3.9 | 4.5 | 5.0 |
| 500 | 4.1 | 5.4 | 5.7 | 4.3 | 4.7 | 5.2 |
| 1000 | 4.3 | 4.9 | 5.1 | 4.5 | 4.9 | 5.1 |
| 2000 | 4.6 | 5.0 | 5.1 | 5.0 | 5.2 | 5.3 |
| 5000 | 5.1 | 5.2 | 5.3 | 5.2 | 5.2 | 5.3 |
| | $O_{ii'} = 1$, $\sum p_{ijk} = 0.60$ $\sum q_{ijk} = 0.40$, $p_{111} = 0.15$ $q_{111} = 0.10$ | | | $O_{ii'} = 1$, $\sum p_{ijk} = 0.80$ $\sum q_{ijk} = 0.20$, $p_{111} = 0.20$ $q_{111} = 0.05$ | | |
| **n** | **Q** | $\chi^2$ | $G^2$ | **Q** | $\chi^2$ | $G^2$ |
| 125 | 2.0 | 4.5 | 6.2 | 2.5 | 8.6 | 13.1 |
| 150 | 2.3 | 4.6 | 6.1 | 2.6 | 6.8 | 9.9 |
| 200 | 3.2 | 4.7 | 5.7 | 2.8 | 5.8 | 7.0 |
| 300 | 3.8 | 5.1 | 5.6 | 3.2 | 4.9 | 5.9 |
| 400 | 3.9 | 4.6 | 5.0 | 3.9 | 4.8 | 5.6 |
| 500 | 4.3 | 4.8 | 5.3 | 4.0 | 5.3 | 5.6 |
| 1000 | 4.5 | 4.8 | 4.9 | 4.4 | 5.0 | 5.2 |
| 2000 | 4.9 | 4.9 | 5.0 | 4.6 | 4.9 | 5.0 |
| 5000 | 5.1 | 5.1 | 5.1 | 5.1 | 5.2 | 5.3 |

*Note*: $Q$: Test based on the odds ratios. $\chi^2$: Pearson test. $G^2$: Likelihood ratio test.

rest of the profiles it is verified that $O_{ii'} = 1$ with $i = 2, \ldots, 7$ and $i' = i + 1, \ldots, 8$. As the nominal error, $\alpha = 5\%$ has been considered.

Table 1 shows the results obtained for some of the scenarios considered. The $Q$ test presents type I error values that increase with the sample size, and very close to the nominal error for $n \geq 500$ or $n \geq 1000$, depending on the scenario. The $Q$ test is somewhat more conservative than $\chi^2$ and $G^2$ tests (mainly for $n \leq 400$), but it does not exceed the nominal error in excess with any sample size in all considered scenarios. The $\chi^2$ test and the $G^2$ test greatly exceed the nominal error in some situations, especially for $n = 125$–$150$ (also for $n = 200$ in the $G^2$ test), and therefore both hypothesis tests can give rise to too many false significances and should not be used for these small sample sizes, as expected. In general, the $\chi^2$ and the $G^2$ test are somewhat less conservative than the $Q$ test for $200 \leq n \leq 400$, and type I errors are very similar for all when $n \geq 500$.

Table 2 shows the results obtained for the powers, indicating the values of the odds ratio of the first profile and the values of the probabilities of the multinomial distribution. No results are shown for $n = 125$–$150$ in two of the scenarios because it has not been possible to generate samples that meet the conditions to apply the chi-square test. In all the hypothesis tests and under the same sample size, the power increases as the values of the odds ratio of the first profile increase.

For the $Q$ test, in general, a sample size between moderate ($n = 125$) and large ($n \geq 500$) is needed for the power to be high (greater than 80%), depending on the scenario. The $\chi^2$ and $G^2$ tests are a little more powerful than the $Q$ test when $n \leq 150$–$200$ (depending on the scenario) because their type I errors are also larger (they can greatly exceed the nominal

**TABLE 2** Powers (in %) of the independence tests.

| | $O_{1i'} = 2$ $\sum p_{ijk} = 0.40 \sum q_{ijk} = 0.60$ $p_{111} = 0.20$ $q_{111} = 0.20$ | | | $O_{1i'} = 3$ $\sum p_{ijk} = 0.80 \sum q_{ijk} = 0.20$ $p_{111} = 0.40$ $q_{111} = 0.05$ | | |
|---|---|---|---|---|---|---|
| ***n*** | ***Q*** | $\chi^2$ | $G^2$ | ***Q*** | $\chi^2$ | $G^2$ |
| 125 | 13.8 | 23.7 | 28.7 | – | – | – |
| 150 | 18.3 | 25.8 | 31.3 | – | – | – |
| 200 | 26.8 | 30.9 | 35.9 | 62.3 | 66.1 | 66.2 |
| 300 | 45.9 | 49.1 | 52.5 | 77.6 | 80.6 | 79.4 |
| 400 | 63.7 | 66.1 | 68.1 | 87.9 | 89.8 | 88.8 |
| 500 | 74.3 | 76.1 | 77.4 | 100 | 100 | 100 |
| 1000 | 98.4 | 98.5 | 98.5 | 100 | 100 | 100 |
| 2000 | 100 | 100 | 100 | 100 | 100 | 100 |
| 5000 | 100 | 100 | 100 | 100 | 100 | 100 |
| | $O_{1i'} = 5$ $\sum p_{ijk} = 0.20 \sum q_{ijk} = 0.80$ $p_{111} = 0.15$ $q_{111} = 0.30$ | | | $O_{1i'} = 15$ $\sum p_{ijk} = 0.60 \sum q_{ijk} = 0.40$ $p_{111} = 0.50$ $q_{111} = 0.10$ | | |
| ***n*** | ***Q*** | $\chi^2$ | $G^2$ | ***Q*** | $\chi^2$ | $G^2$ |
| 125 | – | – | – | 100 | 100 | 100 |
| 150 | – | – | – | 100 | 100 | 100 |
| 200 | 92.6 | 95.1 | 96.4 | 100 | 100 | 100 |
| 300 | 97.9 | 99.2 | 99.4 | 100 | 100 | 100 |
| 400 | 99.6 | 99.8 | 99.9 | 100 | 100 | 100 |
| 500 | 100 | 100 | 100 | 100 | 100 | 100 |
| 1000 | 100 | 100 | 100 | 100 | 100 | 100 |
| 2000 | 100 | 100 | 100 | 100 | 100 | 100 |
| 5000 | 100 | 100 | 100 | 100 | 100 | 100 |

*Note*: $Q$: Test based on the odds ratios. $\chi^2$: Pearson test. $G^2$: Likelihood ratio test.

error) than those of the $Q$ test. In general terms, there is no important difference (less than 1% on average) between the powers of the three methods when $n \geq 300$–400, and it is necessary to have a sample size between moderate ($n = 125$) and large ($n \geq 500$) so that the power is high (over 80%), depending on the values of the odds ratios.

From the results of the Monte Carlo experiments, it follows the $Q$ test has an adequate asymptotic behavior for its practical application: its type I error does not exceed the nominal error and its power is high when the sample size is not excessively large (depending on the value of the odds ratios). The $\chi^2$ and $G^2$ tests should not be applied when $n = 125$–150 (moderate sizes) since their type I errors can exceed the nominal error, giving rise to too many false significances. When the sample size is large, the three methods have a very similar asymptotic behavior.

## 5 | ILLUSTRATIVE EXAMPLE

To illustrate the performance of the model, we have analyzed a data set from Weiner et al.[16] on the diagnosis of coronary disease. In particular, here we are focused on the exercise stress testing (EST) for the diagnosis of coronary disease, using a coronary arteriography as the GS. The proposed model has been implemented in R and the script and data set to reproduce this application are available as supplementary material. Table 3 shows the results obtained when applying the two medical tests to a sample of 2045 people, in which the row profiles are formed by the categories of the variables of sex, Resting ST's & T-Waves (RST-TW) and EST, and in this order. Therefore, the sub-index $r$ of the table refers to sex ($r = 1$ for men and $r = 2$ for women), sub-index $s$ refers to RST-TW ($s = 1$ abnormal and $s = 2$ normal) and sub-index $t$ refers to the result of EST ($t = 1$ positive and $t = 2$ negative).

**TABLE 3** Study of Weiner et al.

| Profile | $(r, s, t)$ | Diseased | Non-diseased | $\alpha_i$ |
|---|---|---|---|---|
| 1 | (1, 1, 1) | 224 | 35 | 1.4022564 |
| 2 | (1, 1, 2) | 32 | 41 | 0.4018787 |
| 3 | (1, 2, 1) | 591 | 80 | 3.6367663 |
| 4 | (1, 2, 2) | 176 | 286 | 2.5196283 |
| 5 | (2, 1, 1) | 59 | 75 | 0.7378854 |
| 6 | (2, 1, 2) | 8 | 43 | 0.2601560 |
| 7 | (2, 2, 1) | 69 | 74 | 0.7911212 |
| 8 | (2, 2, 2) | 33 | 219 | 1.2751462 |
| $\beta_j$ | | 0.954 | 1.05 | $\mu = 175.6074$ |

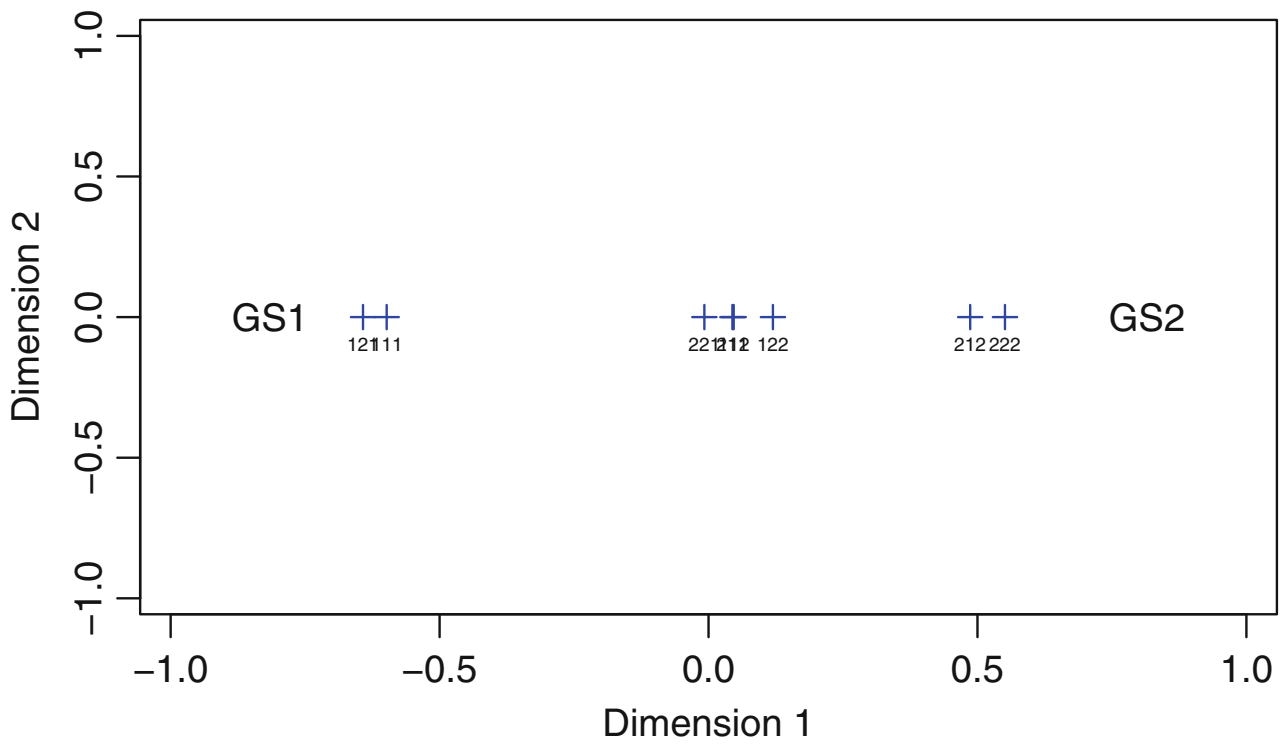

**FIGURE 1** Representation of associations of the profiles of individuals regarding coronary disease.

The estimated expected values when applying the proposed two-step procedure in one dimension, along with the estimated values of the row, column, and overall effects are shown in Table 3. The saturated model perfectly recovers the observed values (3), as is to be expected. Figure 1 shows the representation in one dimension of the associations between the profiles in terms of having or not the disease (this is displayed in two dimensions for easy viewing). As can be appreciated; (1) profiles (1,2,1) (man with normal RST-TW and positive EST) and (1,1,1) (man with abnormal RST-TW and positive EST) are the two profiles most closely associated with having coronary disease; (2) profiles (2,1,2) (woman with abnormal RST-TW and negative EST) and (2,2,2) (woman with normal RST-TW and negative EST) are the profiles most closely associated with not having coronary disease; (3) the rest of the profiles are associated in a similar way with having or not having the disease and also to a similar degree.

The test statistic of the independence test is $Q = 524.5237$ and $P\text{-}value = 0$. If we set $\alpha = 5\%$ then we reject the null hypothesis that all of the log-odds ratios are equal to 0 (ie, that all of the odds ratios are equal to 1). Since the independence test is significant, the unfolding solution for the representation of the associations is not degenerate. Table 4 shows the estimations of the odds ratios between each two profiles and the results of the comparison in pairs (test statistics, $P$-values and adjusted $P$-values). For $\alpha = 5\%$ the adjusted $P$-values using the Holm's method indicate that:

1. The odds of having coronary disease in profile 1 (man with abnormal RST-TW and positive EST) is significantly greater than in profiles 2 (man with abnormal RST-TW and negative EST), 4 (man with normal RST-TW and negative EST), 5 (woman with abnormal RST-TW and positive EST), 6 (woman with abnormal RST-TW and negative EST), 7 (woman with normal RST-TW and positive EST) and 8 (woman with normal RST-TW and negative EST). We do not reject that the odds of having coronary disease in profile 1 is equal to the odds of having the disease in profile 3 (man with normal RST-TW and positive EST). Therefore, a man with positive EST (whatever the status of the RST-TW) has an odds of having coronary disease which are significantly greater than the rest of profiles.

**TABLE 4** Multiple comparisons in the study of Weiner et al.

| Profiles | 2 = (1, 1, 2) | 3 = (1, 2, 1) | 4 = (1, 2, 2) | 5 = (2, 1, 1) | 6 = (2, 1, 2) | 7 = (2, 2, 1) | 8 = (2, 2, 2) |
|---|---|---|---|---|---|---|---|
| 1 = (1, 1, 1) | $\hat{O}_{12} = 8.2$<br>$z_{12} = 7.0659$<br>*P-value* = 0<br>*Adj.P-value* = 0* | $\hat{O}_{13} = 0.8663$<br>$z_{13} = -0.6603$<br>*P-value* = 0.5091<br>*Adj.P-value* = 1 | $\hat{O}_{14} = 10.4$<br>$z_{14} = 11.3978$<br>*P-value* = 0<br>*Adj.P-value* = 0* | $\hat{O}_{15} = 8.1356$<br>$z_{15} = 8.3306$<br>*P-value* = 0<br>*Adj.P-value* = 0* | $\hat{O}_{16} = 34.4$<br>$z_{6} = 8.3095$<br>*P-value* = 0<br>*Adj.P-value* = 0* | $\hat{O}_{17} = 6.8638$<br>$z_{17} = 7.7965$<br>*P-value* = 0<br>*Adj.P-value* = 0* | $\hat{O}_{18} = 42.4727$<br>$z_{18} = 14.3863$<br>*P-value* = 0<br>*Adj.P-value* = 0* |
| 2 = (1, 1, 2) | – | $\hat{O}_{23} = 0.1056$<br>$z_{23} = -8.5054$<br>*P-value* = 0<br>*Adj.P-value* = 0* | $\hat{O}_{24} = 1.2683$<br>$z_{24} = 0.9335$<br>*P-value* = 0.3505<br>*Adj.P-value* = 1 | $\hat{O}_{25} = 0.9921$<br>$z_{25} = -0.00269$<br>*P-value* = 0.9785<br>*Adj.P-value* = 1 | $\hat{O}_{26} = 4.1951$<br>$z_{26} = 3.1756$<br>*P-value* = 0.0015<br>*Adj.P-value* = 0.015* | $\hat{O}_{27} = 0.837$<br>$z_{27} = -0.615$<br>*P-value* = 0.5385<br>*Adj.P-value* = 1 | $\hat{O}_{28} = 5.1796$<br>$z_{28} = 5.467$<br>*P-value* = 0<br>*Adj.P-value* = 0* |
| 3 = (1, 2, 1) | – | – | $\hat{O}_{34} = 12.0047$<br>$z_{34} = 16.2572$<br>*P-value* = 0<br>*Adj.P-value* = 0* | $\hat{O}_{35} = 8.3909$<br>$z_{35} = 10.6204$<br>*P-value* = 0<br>*Adj.P-value* = 0* | $\hat{O}_{36} = 39.7078$<br>$z_{36} = 9.1343$<br>*P-value* = 0<br>*Adj.P-value* = 0* | $\hat{O}_{37} = 7.9228$<br>$z_{37} = 10.0756$<br>*P-value* = 0<br>*Adj.P-value* = 0* | $\hat{O}_{38} = 49.0261$<br>$z_{38} = 17.5729$<br>*P-value* = 0<br>*Adj.P-value* = 0* |
| 4 = (1, 2, 2) | – | – | – | $\hat{O}_{45} = 0.7823$<br>$z_{45} = -1.2361$<br>*P-value* = 0.2164<br>*Adj.P-value* = 1 | $\hat{O}_{46} = 3.3077$<br>$z_{46} = 3.0149$<br>*P-value* = 0.0026<br>*Adj.P-value* = 0.0231* | $\hat{O}_{47} = 0.66$<br>$z_{47} = -2.155$<br>*P-value* = 0.0312<br>*Adj.P-value* = 0.2493 | $\hat{O}_{48} = 4.0839$<br>$z_{48} = 6.7043$<br>*P-value* = 0<br>*Adj.P-value* = 0* |
| 5 = (2, 1, 1) | – | – | – | – | $\hat{O}_{56} = 4.2283$<br>$z_{56} = 3.4123$<br>*P-value* = 0.0006<br>*Adj.P-value* = 0.0071* | $\hat{O}_{57} = 0.8437$<br>$z_{57} = -0.7041$<br>*P-value* = 0.4814<br>*Adj.P-value* = 1 | $\hat{O}_{58} = 5.2306$<br>$z_{58} = 6.4745$<br>*P-value* = 0<br>*Adj.P-value* = 0* |
| 6 = (2, 1, 2) | – | – | – | – | – | $\hat{O}_{67} = 0.1995$<br>$z_{67} = -3.8391$<br>*P-value* = 0.0001<br>*Adj.P-value* = 0.0015* | $\hat{O}_{68} = 1.2347$<br>$z_{68} = 0.4926$<br>*P-value* = 0.6223<br>*Adj.P-value* = 1 |
| 7 = (2, 2, 1) | – | – | – | – | – | – | $\hat{O}_{78} = 6.188$<br>$z_{78} = 7.2686$<br>*P-value* = 0<br>*Adj.P-value* = 0* |

*Test significant to an error $\alpha = 5\%$.

**TABLE 5** Estimations of the parameters.

| Estimators | Covariate patterns (*r*,*s*) Man, abnormal RST-TW (1, 1) | Man, normal RST-TW (1, 2) | Woman, abnormal RST-TW (2, 1) | Woman, normal RST-TW (2, 2) |
|---|---|---|---|---|
| $\hat{\psi}_{rs}$ | 0.7711 | 0.6770 | 0.3622 | 0.2582 |
| $\widehat{Se}_{rs}$ | 0.8750 | 0.7705 | 0.8806 | 0.6765 |
| $\widehat{Sp}_{rs}$ | 0.5395 | 0.7814 | 0.3644 | 0.7474 |
| $\widehat{LR}_{rs}^{+}$ | 1.9 | 3.5252 | 1.3855 | 2.6785 |
| $\widehat{LR}_{rs}^{-}$ | 0.2317 | 0.2937 | 0.3277 | 0.4328 |
| $\widehat{PPV}_{rs}$ | 0.8649 | 0.8808 | 0.4403 | 0.4825 |
| $\widehat{NPV}_{rs}$ | 0.5616 | 0.6190 | 0.8431 | 0.8690 |
| $\widehat{PPV}_{rs}/\left(1-\widehat{PPV}_{rs}\right)$ | 6.4 | 7.3875 | 0.7867 | 0.9324 |
| $\widehat{NPV}_{rs}/\left(1-\widehat{NPV}_{rs}\right)$ | 0.7805 | 0.6154 | 0.1860 | 0.1507 |

2. The odds of having the disease in profile 2 (man with abnormal RST-TW and negative EST) is significantly lower than the odds of having the disease in profiles 3 (man with normal RST-TW and positive EST), and significantly higher than in profiles 6 (woman with abnormal RST-TW and negative EST) and 8 (woman with abnormal RST-TW and negative EST). We do not reject that the odds of having coronary disease in profile 2 is equal to the odds of having the disease in profiles 4 (man with normal RST-TW and negative EST), 5 (woman with an abnormal RST-TW and positive EST) and 7 (woman with normal RST-TW and positive EST).
3. The odds of having the disease in profile 3 are significantly higher than the odds of having the disease in profiles 2, 4, 5, 6, 7 and 8.
4. The odds of having the disease in profile 4 are significantly higher than the odds of having the disease in profiles 6 and 8. We do not reject that the odds of having coronary disease in profile 4 is equal to the odds of having the disease in profiles 5 and 7.
5. The odds of having the disease in profile 5 is significantly higher than the odds of having the disease in profiles 6 and 8. Therefore, a woman with abnormal RST-TW and positive EST has a greater odds of having coronary disease than a woman with negative EST (whether or not the RST-TW is normal). We do not reject that the odds of having the disease in profile 5 is equal to the odds of having the disease in profile 7, and therefore we do not reject that a woman with positive EST has the same odds of having coronary disease whatever the status of the RST-TW.
6. The odds of having the disease in profile 6 is significantly lower than the odds of having the disease in profile 7, and therefore a woman with abnormal RST-TW and negative test has a lower odds than a woman with normal RST-TW and positive EST. We do not reject that the odds of having coronary disease is equal among profiles 6 and 8, and therefore we do not reject that a woman with negative EST has the same odds of having coronary disease whatever the status of the RST-TW.
7. The odds of having the disease in profile 7 are significantly higher than the odds of having the disease in profile 8, and therefore a woman with normal RST-TW and positive EST has a greater odds of having the disease than a woman with normal RST-TW and negative EST.

Therefore, the results obtained lead to the following conclusions:

1. Profiles 1 and 3 (non-significant test) both have greater odds of having the disease than the rest of the profiles, and therefore they are the profiles with the greatest disease of coronary disease.
2. Profiles 6 and 8 (non-significant test) are those with the lowest odds of having the disease, and therefore they are the profiles with the lowest risk of coronary disease.
3. For profiles 2, 4, 5, and 7, the individual tests between them are not significant and, therefore, there is no evidence of any differences regarding their positions in the unfolding representation and thus of their associations with the coronary disease.

Regarding the disease prevalence and the parameters of the EST, Table 5 shows their estimations in the different patterns of the covariates. Applying the equations of appendix D we obtain the estimations of the odds ratios given in Table 4. For those profiles which are more associated with having the disease, it is obtained that: (a) the odds of a man with an abnormal RST-TW and a positive EST having the disease is 6.4, and (b) the odds of a man with a normal RST-TW and a positive EST having the disease is approximately 7.4. In both cases, the odds are calculated as $\widehat{PPV}_{rs}/\left(1-\widehat{PPV}_{rs}\right)$ since the result of the BDT is positive in the two profiles. The rest of the odds are interpreted in a similar way.

## 6 | DISCUSSION

In this paper, a two-step log-linear procedure to estimate and represent associations in full dimension is proposed. The log-linear model is shown to be equivalent to a distance association model in full dimension, after an appropriate parameterization, which enables the direct representation of associations between rows and columns in a two-way contingency table resulting from cross-classified data sets.

When one of the two variables involved in the contingency table is binary, as is usually the case in the diagnosis of a disease, the relation between independency and equality of localization of the points in the unfolding representation is shown. Therefore, the interpretation of close positions in the representation can be made in terms of equal degrees of association from a statistical point of view. This also enables the statistical identification of a degenerate solution in unfolding in terms of the odds ratio.

In general, the low-dimensional distance association model is not equivalent to the related log-linear model. Therefore, a procedure is considered for testing independence in terms of the observed frequencies, these being the expected values estimated by the DA model in full dimension, which is based on testing that the values of the odds ratio are all equal to one. The test statistic $Q$ is distributed asymptotically according to a chi-squared distribution with $I-1$ degrees of freedom, since only $I-1$ odds ratio must be considered, and its performance is compared with that of classical independence tests. From the results obtained in the simulation experiments, it can be appreciated that, in general, the proposed $Q$ statistic has a good asymptotic behavior, its performance being better than the classical tests for moderate samples.

Consideration of the natural logarithmic transformation of the odds ratio for the test statistic ensures its invariance in the face of a change in the baseline profile. Although other transformations have been proposed in this framework, such as the inverse hyperbolic sine or the inverse sine,[17,18] it is easy to show that these are not invariant for a change in the baseline profile, which is a clear drawback.

To determine the causes of a significant test, individual hypothesis tests on each odds ratio have been proposed using Holm's method to adjust the P-values obtained. The situation of independence is not precisely what is of interest here, since we are mainly focused on the analysis of associations. Indeed, to analyze global independence, the classic tests based on the chi-square distribution (Pearson and likelihood ratio) can also be used, as well as Fisher's exact test, all of them based on the observed frequencies since the model is saturated. Nevertheless, there are certain aspects that should be highlighted in this regard. The hypothesis tests based on the chi-square distribution, that is, $\chi^2$ test and $G^2$ test, require certain conditions on the expected frequencies in order to be applied, which means that their use is somewhat limited for small or even moderate samples (see, for instance, Cochran[14] and McDonald[15]). On the other hand, when the test is significant, the investigation of the causes of significance is done by partitioning the table, which is computationally expensive as the number of profiles increases. Although partitioning the table of observed frequencies into $I-1$ subtables, the test statistic $G^2$ is the sum of the $I-1$ test statistics $G^2$, this property does not hold for the $Q$ statistics, nor is it true for the $\chi^2$ statistics.

The same occurs with Fisher's exact test, which may even be computationally unfeasible, as well as the subsequent investigation of the causes of the lack of independence. The global test based on the odds ratio has advantages over the previous tests in this framework: (a) it can always be applied (for zero frequencies it is enough to simply add 0.5), and (b) if the test is significant, the causes of the significance are investigated simple and fast, since all the parameters involved have been estimated when solving the global test.

In particular, the clinical situation in which the profiles are composed of two binary covariates and a BDT has been widely described, given its great interest in medical research. In this situation, the relation of the odds ratio for two profiles with measures of quality of the BDT in the different patterns of covariates is set. In addition, simulation experiments have been carried out to study the asymptotic behavior of the independence test for the case of eight profiles, and the hypothesis test showed good performance both in terms of type I error and power.

An interesting topic to investigate is the relationship between equality of odds ratio and location of points when the model is estimated in low dimension. This is particularly interesting for tables in which the large number of profiles induces the use of a combined latent class distance association model such as the LCDA.[3] In addition, given the relationship with the RC(M) model, the study of this methodology in the RC(M) framework in low dimension, and also in related models collapsing categories such as the Kateri and Iliopuolos model,[19] is of great interest.

## FUNDING INFORMATION

This work has been partially supported by the Ministry of Science and Innovation-State Research Agency/10.13039/501100011033/Spain, by "ERDF A way of making Europe" [grant number RTI2018-099723-B-I00]; grant B-CTS-184-UGR20 funded by ERDF, EU/Ministry of Economic Transformation, Industry, Knowledge and Universities of Andalusia (J. Fernando Vera); Spanish Ministry of Science and Innovation, Grant PID2021-126095NB-100 funded by MCIN/AEI/10.13039/501100011033 and by "ERDF A way of making Europe" (J. A. Roldán-Nofuentes).

## CONFLICT OF INTEREST STATEMENT

The authors declare no potential conflict of interests.

## DATA AVAILABILITY STATEMENT

Data sharing is not applicable to this article as no new data were created or analyzed in this study.

## ORCID

*J. Fernando Vera* https://orcid.org/0000-0002-6499-7132
*José A. Roldán-Nofuentes* https://orcid.org/0000-0003-0251-5588

## SUPPORTING INFORMATION

Additional supporting information can be found online in the Supporting Information section at the end of this article.

## APPENDIX A

The logarithm of the estimated expected frequencies, $\log(\mu_{ij})$, in a log-linear model can be written as

$$\log(\mu_{ij}) = \widetilde{\lambda} + \widetilde{\lambda}_i^R + \widetilde{\lambda}_j^C + \widetilde{\lambda}_{ij}^{RC}, \tag{A1}$$

where $\widetilde{\lambda}$ is the mean of the logarithm of the expected frequencies, $\widetilde{\lambda}_i^R$ and $\widetilde{\lambda}_j^C$ are the row and column deviation values with respect to the mean, respectively, and $\widetilde{\lambda}_{ij}^{RC}$ is the logarithm of the estimated interaction effect (see, eg, Everit[20]). Hence, considering the singular value decomposition of the matrix of interaction values we can write $\widehat{\boldsymbol{\lambda}}^{RC} = \mathbf{U\Gamma\Lambda}^t = 2\mathbf{XY}^t$, and therefore, (A1) can be written as follows:

$$\log(\mu_{ij}) = \widetilde{\lambda} + \left(\widetilde{\lambda}_i^R + \mathbf{x}_i'\mathbf{x}_i\right) + \left(\widetilde{\lambda}_j^C + \mathbf{y}_j'\mathbf{y}_j\right) - d^2(\mathbf{x}_i, \mathbf{y}_j),$$

where $\mathbf{x_i}$ and $\mathbf{y_j}$ are the $i$th row of $\mathbf{X}$ and the $j$th row of $\mathbf{Y}$, respectively, in full dimension. Hence, we have in (2) that $\mu = \exp(\widetilde{\lambda})$, $\alpha = \exp\left(\widetilde{\lambda}_i^R - \mathbf{x}_i'\mathbf{x}_i\right)$, and $\beta = \exp\left(\widetilde{\lambda}_j^C - \mathbf{y}_j'\mathbf{y}_j\right)$, and we will assume the association parameters $\theta_{ij}$ can be expressed in full dimension in terms of the Euclidean distances $d_{ij}^2 = d^2(\mathbf{x}_i, \mathbf{y}_j)$,

$$d^2(\mathbf{x}_i, \mathbf{y}_j) = \sum_{m=1}^{M} (x_{im} - y_{jm})^2,$$

using the exponential form $\theta_{ij} = \exp\left(-d_{ij}^2\right)$. Thus, in full dimension, taking the logarithm of (2) we can express the model in log-linear terms as follows,

$$\log(\mu_{ij}) = \lambda + \lambda_i + \lambda_j - d_{ij}^2(\mathbf{x}_i, \mathbf{y}_j),$$

where $\log(\theta_{ij}) = -d_{ij}^2$.

## APPENDIX B

For cross classified data involving a dichotomous variable in the two-way table, the unfolding representation is in one dimension. Then, denoting by $J$ and $j$ the two column categories in the table, it follows that for any $i, i' = 1, \dots, I$ row categories,

$$O_{ij} = O_{i'j} \Leftrightarrow d_{iJ} - d_{ij} = d_{i'J} - d_{i'j} \tag{B1}$$

which only occurs when the distance between both row categories $i$ and $i'$ is zero. This is true since, $O_{ij} = O_{i'j}$ if and only if $(d_{iJ} - d_{ij})(d_{iJ} + d_{ij}) = (d_{i'J} - d_{i'j})(d_{i'J} + d_{i'j})$. Thus, if the solution is not degenerate ($d_{jJ} \neq 0$) and $O_{ij} = O_{i'j}$ for any two row categories $i$ and $i'$, one of the following three situations can occur between the location of these four points on a straight line:

- The two column categories are located at the ends of the graph, for example, $J, i, i', j$. Then, $\left(d_{iJ} + d_{ij}\right) = \left(d_{i'J} + d_{i'j}\right) = d_{jJ}$, and therefore $d_{iJ} - d_{i'J} = d_{ij} - d_{i'j}$, that is, $-d_{ii'} = d_{ii'}$, if and only if $d_{ii'} = 0$.
- There is a column category located between the two row categories, for example $i, j, i', J$. Then, $d_{iJ} - d_{i'J} = d_{ij} + d_{i'j} = d_{ii'}$. If we suppose that $d_{ii'} \neq 0$, then $d_{iJ} - d_{ij} = d_{i'j} - d_{i'J}$, that is, $d_{iJ} = -d_{iJ}$, which is a contradiction. Hence, $d_{ii'} = 0$.
- Row categories are placed together, as are column categories, for example, $i, i', j, J$. Then, $d_{iJ} - d_{ij} = d_{i'J} - d_{i'j} = d_{jJ}$, and therefore, $d_{iJ} - d_{i'J} = d_{i'j} - d_{ij}$, that is, $d_{ii'} = -d_{ii'}$, if and only if $d_{ii'} = 0$.

## APPENDIX C

To simplify the demonstration, let us suppose that the global hypothesis test is solved taking profile 1 as the baseline profile, then $\hat{\mathbf{U}}_1 = \left(\hat{U}_{12}, \hat{U}_{13}, \ldots, \hat{U}_{1I}\right)^T$ and the test statistic is

$$Q_1 = \hat{\mathbf{U}}_1^T \widehat{\sum}_{\hat{\mathbf{U}}_1}^{-1} \hat{\mathbf{U}}_1.$$

Let us now consider that the global hypothesis test is solved taking profile 2 as the baseline profile. As the odds ratios verify that $O_{ii} = 1$ and that $O_{ii'} = O_{i''i'}/O_{i''i}$, with $i' \neq i''$, then $U_{ii} = 0$ and $U_{ii'} = \log(O_{ii'}) = \log(O_{i''i'}/O_{i''i}) = U_{i''i'} - U_{i''i}$. Applying these properties, vector $\mathbf{U}_2$ is written in terms of the components of vector $\mathbf{U}_1$ as

$$\mathbf{U}_2 = (U_{21}, U_{23}, \ldots, U_{2I})^T = (-U_{12}, U_{13} - U_{12}, \ldots, U_{1I} - U_{12})^T.$$

Then, the variance-covariance matrix of $\hat{\mathbf{U}}_2$ can be estimated from the variance-covariance matrix of $\hat{\mathbf{U}}_1$ applying the delta method, that is,

$$\widehat{\sum}_{\hat{\mathbf{U}}_2} = \left(\frac{\partial \mathbf{U}_2}{\partial \mathbf{U}_1}\right) \widehat{\sum}_{\hat{\mathbf{U}}_1} \left(\frac{\partial \mathbf{U}_2}{\partial \mathbf{U}_1}\right)^T,$$

where $\left(\frac{\partial \mathbf{U}_2}{\partial \mathbf{U}_1}\right)$ is a matrix of a dimension $(I-1) \times (I-1)$ whose elements are constant, that is,

$$\frac{\partial \mathbf{U}_2}{\partial \mathbf{U}_1} = \begin{pmatrix} -1 & 0 & 0 & 0 & \ldots & 0 \\ -1 & 1 & 0 & 0 & \ldots & 0 \\ -1 & 0 & 1 & 0 & \ldots & 0 \\ \vdots & \vdots & \vdots & \vdots & \vdots & \vdots \\ -1 & 0 & 0 & 0 & \ldots & 1 \end{pmatrix}.$$

In this matrix, the elements of the first column are equal to $-1$, the rest of the elements in the main diagonal are equal to 1 and all of the other elements of the matrix are equal to 0. It is easy to verify that

$$\frac{\partial \mathbf{U}_2}{\partial \mathbf{U}_1} = \left(\frac{\partial \mathbf{U}_2}{\partial \mathbf{U}_1}\right)^{-1}$$

and that

$$\left(\frac{\partial \mathbf{U}_2}{\partial \mathbf{U}_1}\right)^{-1} \hat{\mathbf{U}}_2 = \hat{\mathbf{U}}_1.$$

Then the test statistic for the global test is

$$Q_2 = \hat{\mathbf{U}}_2^T \widehat{\sum}_{\hat{\mathbf{U}}_2}^{-1} \hat{\mathbf{U}}_2 = \hat{\mathbf{U}}_2^T \left[\left(\frac{\partial \mathbf{U}_2}{\partial \mathbf{U}_1}\right) \widehat{\sum}_{\hat{\mathbf{U}}_1} \left(\frac{\partial \mathbf{U}_2}{\partial \mathbf{U}_1}\right)^T\right]^{-1} \hat{\mathbf{U}}_2$$

$$= \hat{\mathbf{U}}_2^T \left[ \left( \frac{\partial \mathbf{U}_2}{\partial \mathbf{U}_1} \right)^{-1} \right]^T \widehat{\sum}_{\hat{\mathbf{U}}_1}^{-1} \left( \frac{\partial \mathbf{U}_2}{\partial \mathbf{U}_1} \right)^{-1} \hat{\mathbf{U}}_2$$
$$= \hat{\mathbf{U}}_1^T \widehat{\sum}_{\hat{\mathbf{U}}_1}^{-1} \hat{\mathbf{U}}_1 = Q_1.$$

The demonstration is similar for the rest of profiles $i \geq 3$. In general, for any other profiles $i$ and $i'$, it holds that.

$$\mathbf{U}_i = (U_{ii''})^T, \text{with } i'' = 1, \ldots, I \text{ and } i'' \neq i,$$

and

$$\mathbf{U}_{i'} = (U_{i'i''})^T, \text{with } i'' = 1, \ldots, I \text{ and } i'' \neq i'.$$

The elements of $\mathbf{U}_{i'}$ are written in terms of the elements of $\mathbf{U}_i$ as.

$$\hat{U}_{i'i''} = \begin{cases} -\hat{U}_{ii'}, \mathrm{i} = i'' \\ \hat{U}_{ii''} - \hat{U}_{ii'}, \mathrm{i} \neq i'' \end{cases}, \text{with } i'' = 1, \ldots, I.$$

The matrix of partial derivatives $\partial \mathbf{U}_{i'} / \partial \mathbf{U}_i$ has the following elements: the elements of column $i$ are all equal to $-1$, the rest of the elements in the main diagonal are all equal to 1, and the rest of the elements in the matrix are all equal to 0. A matrix of this type always verifies that $\partial \mathbf{U}_{i'} / \partial \mathbf{U}_i = (\partial \mathbf{U}_{i'} / \partial \mathbf{U}_i)^{-1}$ and that $(\partial \mathbf{U}_{i'} / \partial \mathbf{U}_i)^{-1} \mathbf{U}_{i'} = \mathbf{U}_i$.

## APPENDIX D

In clinical practice, it is common for the diagnosis of a disease to be made based on the result of a BDT and the observation of binary covariates (eg, sex, family history, the presence of a risk factor, etc.). Here, we will consider that two binary covariates are observed, although extending this to more than two covariates is simple. Moreover, the extension to two (or more) BDTs is also simple, and in this case it is necessary to consider the covariance between the two (or more than two) BDTs when calculating the probabilities of the profiles. In the situation of a single BDT and two binary covariates (and also with a single covariate), the odds ratio between two profiles is written, as detailed further on, in terms of measures of the quality of the BDT in each pattern of covariates (or of the single covariate in such a situation).

Let us consider that for all of the $n$ individuals in a random sample a gold standard and a BDT are applied, and that for all of these individuals we observe two binary covariates. This situation leads to the profiles and the frequencies given in Table D1. The theoretical probabilities are defined as $p_{rst} = P(GS = 1, A_1 = r, A_2 = s, T = t)$ and $q_{rst} = P(GS = 2, A_1 = r, A_2 = s, T = t)$, with $r, s, t = 1, 2$ and verifying that $\sum_{r,s,t=1}^{2} p_{rst} + \sum_{r,s,t=1}^{2} q_{rst} = 1$. These probabilities are expressed in terms of the sensitivity and the specificity of the BDT as

$$p_{rs1} = P(GS = 1, A_1 = r, A_2 = s, T = 1) = \tau_{rs}\psi_{rs}Se_{rs},$$

$$p_{rs2} = P(GS = 1, A_1 = r, A_2 = s, T = 2) = \tau_{rs}\psi_{rs}(1 - Se_{rs}),$$

$$q_{rs1} = P(GS = 2, A_1 = r, \mathrm{A}_2 = r, T = 1) = \tau_{rs}(1 - \psi_{rs})\left(1 - Sp_{rs}\right)$$

and

$$q_{rs2} = P(GS = 2, A_1 = r, \mathrm{A}_2 = r, T = 2) = \tau_{rs}(1 - \psi_{rs})Sp_{rs},$$

where $\tau_{rs} = P(A_1 = r, A_2 = s)$, $\psi_{rs} = P(GS = 1|A_1 = r, A_2 = s)$, $Se_{rs} = P(T = 1|GS = 1, A_1 = r, A_2 = s)$ is the sensitivity of the BDT for individuals with $A_1 = r$ and $A_2 = s$, and $Sp_{rs} = P(T = 2|GS = 2, A_1 = r, A_2 = s)$ is the specificity of the BDT for individuals with $A_1 = r$ and $A_2 = s$. The maximum likelihood estimators of $p_{rst}$ and $q_{rst}$ are $\hat{p}_{rst} = a_{rst}/n$ and $\hat{q}_{rst} = b_{rst}/n$, where $n = \sum_{\mathrm{r,s,t}=1}^{2} a_{rst} + \sum_{r,s,t=1}^{2} b_{rst}$, and the maximum likelihood estimators of the disease prevalence and of the sensitivity and the specificity of the BDT in the different patterns of covariates are

$$\hat{\psi}_{ij} = \frac{a_{rs1} + a_{rs0}}{a_{rs1} + b_{rs1} + a_{rs0} + b_{rs0}},$$

$$\hat{S}e_{rs} = \frac{a_{rs1}}{a_{rs1} + a_{rs0}} \quad \text{and} \quad \hat{S}p_{rs} = \frac{b_{rs0}}{b_{rs0} + b_{rs1}}.$$

**TABLE D1** Profiles and observed frequencies when $I = 8$.

| Profile | $(r, s, t)$ | Diseased ($GS = 1$) | Non-diseased ($GS = 2$) |
|---|---|---|---|
| 1 | (1, 1, 1) | $a_{111}$ | $b_{111}$ |
| 2 | (1, 1, 2) | $a_{112}$ | $b_{112}$ |
| 3 | (1, 2, 1) | $a_{121}$ | $b_{121}$ |
| 4 | (1, 2, 2) | $a_{122}$ | $b_{122}$ |
| 5 | (2, 1, 1) | $a_{211}$ | $b_{211}$ |
| 6 | (2, 1, 2) | $a_{212}$ | $b_{212}$ |
| 7 | (2, 2, 1) | $a_{221}$ | $b_{221}$ |
| 8 | (2, 2, 2) | $a_{222}$ | $b_{222}$ |

The above 16 probabilities are arranged in vector form as follows:

$$\boldsymbol{\pi} = (p_{111}, p_{112}, \ldots, q_{221}, q_{222})^T,$$

and the odds ratio between profile $i = (r, s, t)$ and profile $i' = \left(r', s', t'\right)$ is expressed as follows:

$$O_{ii'} = \frac{p_{rst}/q_{rst}}{p_{r's't'}/q_{r's't'}}.$$

The odds that profile $i = (r, s, t)$ has the disease when the BDT is positive is

$$\frac{p_{rst}}{q_{rst}} = \frac{\psi_{rs}}{1 - \psi_{rs}} \times \frac{Se_{rs}}{1 - Sp_{rs}} = \frac{\psi_{rs}}{1 - \psi_{rs}} \times LR_{rs}^{+},$$

and the odds that profile $i = (r, s, t)$ has the disease when the BDT is negative is

$$\frac{p_{rst}}{q_{rst}} = \frac{\psi_{rs}}{1 - \psi_{rs}} \times \frac{1 - Se_{rs}}{Sp_{rs}} = \frac{\psi_{rs}}{1 - \psi_{rs}} \times LR_{rs}^{-},$$

where $LR_{rs}^{+} = \frac{Se_{rs}}{1-Sp_{rs}}$ is the positive likelihood ration of the BDT in the pattern of covariates $A_1 = r$ and $A_2 = s$, and $LR_{rs}^{-} = \frac{1-Se_{rs}}{Sp_{rs}}$ is the negative likelihood ratio is the negative likelihood ratio of the BDT in the pattern of covariates $A_1 = r$ and $A_2 = s$. The product $\frac{\psi_{rs}}{1-\psi_{rs}} \times LR_{rs}^{+}$ is the post-test odds when the BDT is positive in the pattern of covariates $A_1 = r$ and $A_2 = s$, that is,

$$\frac{\psi_{rs}}{1 - \psi_{rs}} \times LR_{rs}^{+} = \frac{PPV_{rs}}{1 - PPV_{rs}},$$

where $PPV_{rs}$ is the positive predictive value of the BDT when $A_1 = r$ and $A_2 = s$. In an analogous way, the product $\frac{\psi_{rs}}{1-\psi_{rs}} \times LR_{rs}^{-}$ is the post-test odds when the BDT is negative and $A_1 = r$ and $A_2 = s$, that is,

$$\frac{\psi_{rs}}{1 - \psi_{rs}} \times LR_{rs}^{-} = \frac{1 - NPV_{rs}}{NPV_{rs}},$$

where $NPV_{rs}$ is the negative predictive value of the BDT when $A_1 = r$ and $A_2 = s$. Based on these expressions, the odds ratio between profile $i$ and profile $i'$ is written in terms of the parameters of performance of the BDT as follows:

$$O_{ii'} = \begin{cases} \frac{LR^+_{rs}}{LR^-_{rs}}, r = r', s = s', t = 1, t' = 2 \\ \frac{LR^-_{rs}}{LR^+_{rs}}, r = r', s = s', t = 2, t' = 1 \\ \frac{\psi_{rs}}{1-\psi_{rs}} LR^+_{rs} \times \frac{1-\psi_{rs'}}{\psi_{rs'}} \frac{1}{LR^-_{rs'}}, r = r', s \neq s', t = 1, t' = 2 \\ \frac{\psi_{rs}}{1-\psi_{rs}} LR^-_{rs} \times \frac{1-\psi_{rs'}}{\psi_{rs'}} \frac{1}{LR^+_{rs'}}, r = r', s \neq s', t = 2, t' = 1 \\ \frac{\psi_{rs}}{1-\psi_{rs}} LR^+_{rs} \times \frac{1-\psi_{rs'}}{\psi_{rs'}} \frac{1}{LR^+_{rs'}}, r = r', s \neq s', t = t' = 1 \\ \frac{\psi_{rs}}{1-\psi_{rs}} LR^-_{rs} \times \frac{1-\psi_{rs'}}{\psi_{rs'}} \frac{1}{LR^-_{rs'}}, r = r', s \neq s', t = t' = 2 \\ \frac{\psi_{rs}}{1-\psi_{rs}} LR^+_{rs} \times \frac{1-\psi_{r's}}{\psi_{r's}} \frac{1}{LR^-_{r's}}, r \neq r', s = s', t = 1, t' = 2 \\ \frac{\psi_{rs}}{1-\psi_{rs}} LR^-_{rs} \times \frac{1-\psi_{r's}}{\psi_{r's}} \frac{1}{LR^+_{r's}}, r \neq r', s = s', t = 2, t' = 1 \\ \frac{\psi_{rs}}{1-\psi_{rs}} LR^+_{rs} \times \frac{1-\psi_{r's}}{\psi_{r's}} \frac{1}{LR^+_{r's}}, r \neq r', s = s', t = t' = 1 \\ \frac{\psi_{rs}}{1-\psi_{rs}} LR^-_{rs} \times \frac{1-\psi_{r's}}{\psi_{r's}} \frac{1}{LR^-_{r's}}, r \neq r', s = s', t = t' = 2 \end{cases}.$$

The quotient $\frac{LR^+_{rs}}{LR^-_{rs}} = \frac{Se_{rs}Sp_{rs}}{(1-Sp_{rs})(1-Se_{rs})}$, called the diagnostic odds ratio, is the odds ratio between the BDT and gold standard when $A_1$ ($A_2$) has the same value in both profiles and the BDT is positive in the first profile and negative in the second. Furthermore, when $r \neq r'$ and/or $s \neq s'$, the odds ratio between profile $i$ and profile $i'$ is equal to the product of the post-test odds (or of its opposite) of the BDT in each profile.

Regarding the variance-covariance matrix, applying the delta method it is obtained that

$$\widehat{V}ar\left(\widehat{U}_{ii'}\right) = \widehat{V}ar\left(\widehat{U}_{i'i}\right) = \frac{1}{a_{rst}} + \frac{1}{b_{rst}} + \frac{1}{a_{r's't'}} + \frac{1}{b_{r's't'}}, \text{ with } i, i' = 1, \ldots, I, \; i < i',$$

and

$$\widehat{C}ov\left(\widehat{U}_{ii'}, \widehat{U}_{ii''}\right) = \frac{1}{a_{rst}} + \frac{1}{b_{rst}}, \text{ with } i' \neq i''$$

Expression of $\widehat{V}ar\left(\widehat{U}_{ii'}\right)$ is equal to the one given in Agresti.[9] Hypothesis test (12) is then solved as shown above. Since the test statistic has a long and complicated expression, we use the R statistical software to calculate it. If the independence test is significant to $\alpha$ error, then the causes of the significance are investigated solving the 28 individual tests ($I = 8$) and the Holm's method is applied.